This article is a corrected version of the original one published in Inclusive Smart Cities and e-Health (ICOST2015) Lecture Notes in Computer Science (LNCS) Volume 9102, 2015, pp 181-193. We would like to thank the LNCS Springer for their Archiving Policies for Open Access Repositories. The original article can be found here: http://link.springer.com/chapter/10.1007%2F978-3-319-19312-0_15 The original version missed one reference and thus resulted in errors in references.

# Anti-Fall: A Non-intrusive and Real-time Fall Detector Leveraging CSI from Commodity WiFi Devices

Daqing Zhang[1, 2],Hao Wang[1,2], Yasha Wang[1, 3], Junyi Ma[1,2]

1 Key Laboratory of High Confidence Software Technologies, Ministry of Education, Beijing 100871, China
2 School of Electronics Engineering and Computer Science, Peking University, China
3 National Engineering Research Center of Software Engineering, Peking University, China
{dqzhang, wanghao13, wangys, majy}@sei.pku.edu.cn

**Abstract.** Fall is one of the major health threats and obstacles to independent living for elders, timely and reliable fall detection is crucial for mitigating the effects of falls. In this paper, leveraging the fine-grained Channel State Information (CSI) and multi-antenna setting in commodity WiFi devices, we design and implement a real-time, non-intrusive, and low-cost indoor fall detector, called Anti-Fall. For the first time, the CSI phase difference over two antennas is identified as the salient feature to reliably segment the fall and fall-like activities, both phase and amplitude information of CSI is then exploited to accurately separate the fall from other fall-like activities. Experimental results in two indoor scenarios demonstrate that Anti-Fall consistently outperforms the state-of-the-art approach WiFall, with 10% higher detection rate and 10% less false alarm rate on average.

## 1 Introduction

Falls are the leading cause of fatal and nonfatal injuries to elders in the modern society [1]. In 2010 falls among older adults cost the U.S. health care system over $30 billion dollars [2]. According to the Centers for Disease Control and Prevention, one out of three adults aged 65 and over falls each year [2]. Most elderly people are unable to get up by themselves after a fall, studies have shown that the medical outcome of a fall is largely dependent on the response and rescue time [3], thus timely and automatic detection of falls has long been the research goal in the assistive living community.

Various techniques ranging from ambient device-based to wearable sensor-based solutions have been proposed and studied [3][4][5]. As the most popular ambient device-based solution, the vision-based fall detection systems require high-resolution cameras installed and a series of images recorded for scene recognition, the main problem is the privacy intrusion and inherent requirement for line of sight and lighting condition [5]. The other ambient device-based fall detection systems [3] [4] try to make use of ambient information, e.g., audio noise or floor vibration, caused by falls to detect the risky activity. The main problem with these systems is the high cost incurred and the false alarm caused by other sources leading to similar audio noise and floor vibration as human falls. Both wearable sensor-based [6] and smartphone based [7] fall detection systems employ sensors like accelerators to sense the acceleration or velocity on three axis. However, carrying wearable devices or smartphones are not always possible in home environment, especially for elders.

Due to the limitations of the above-mentioned fall detection solutions, very few fall detection systems have been widely deployed in real home settings so far [8]. In recent years, the rapid development in wireless techniques has stimulated the research in studying the relationship between the wireless signal and human activities. In particular, the recently exposed physical layer **Channel State Information (CSI)** on commercial WiFi devices reveals multipath channel features **at the granularity of OFDM subcarriers [9],** which is much finer-grained than the traditional MAC layer RSS**.** Significant progress has been made in applications in motion detection [10][13], gesture recognition and activity recognition [11] [12]. The rationale behind all these research efforts is that wireless signals are affected in a different way by different human activities, and human activities can be recognized in real-time by understanding the wireless signal patterns.

With this motivation, in this paper, we propose a real-time, non-intrusive and robust fall detection system, called Anti-Fall, leveraging cheap and widely deployed WiFi devices at home, without requiring the subjects to wear or carry any objects. The main contributions of this work are as follows:

1. To the best of our knowledge, we are the first to use **both the phase and amplitude features of CSI** in WiFi to detect falls in real-time in indoor environments. Anti-Fall proves to be the first effective and automatic **activity segmentation and fall detection system** using commodity WiFi devices.
2. Instead of collecting the fall and other activity RF signals manually for training and testing, we find a robust way to **segment the fall and fall-like activities** using the **phase difference of CSI over two antennas** as the salient feature. We further extract features from the amplitude and phase information of CSI, which harness both the space and frequency diversity, to differentiate the fall and fall-like activities.
3. We prototype Anti-Fall on commodity WiFi devices and validate its performance in different indoor environments. Experiment results demonstrate that Anti-Fall can segment the fall and fall-like activities reliably in the WiFi wireless signal streams and consistently outperform the state-of-the-art fall detector WiFall, with fall detection precision of 89% and false alarm rate of 13% on average.

The rest of the paper is organized as follows. We first review the related work in Section 2. Then we introduce some preliminaries about channel state information and fall activities targeted in Section 3. In Section 4, we present the detailed system design and algorithms of our proposed fall detector, Anti-Fall. Followed by the evaluation and comparison results in Section 5. Finally, we conclude the work in Section 6.

## 2 Related Work

In this section, we review the related work from two perspectives: *research on fall detection* and *research on WiFi CSI-based activity recognition*.

**Related Work on Fall Detection.** Fall detection has attracted a lot of attention in assistive living and healthcare community in the past two decades. Yu [5] and Natthapon et al. [8] reviewed the principles and approaches used in existing fall detection systems. Roughly, the fall detection systems can be classified into two broad categories: ambient device based systems and wearable sensor based systems. *The ambient device based fall detection systems* intend to detect falls in a non-intrusive way by exploiting the ambient information including visual [5], audio [4] and floor vibration [3] data produced by a fall. The earliest and most researched approach in this category is based on vision techniques. In these systems, high resolution cameras are equipped in the monitoring environment and a series of images are recorded. By using activity classification algorithm, the fall activity is distinguished from other events [5]. The major problem with vision-based methods is the privacy intrusion, especially in the bathroom setting. Besides, the vision-based fall detection systems fail to work in darkness or when the elders stay outside of the focus of the cameras. The other type of ambient device based fall detection systems are based on the principle that different human activities will cause different changes in acoustic noise or floor vibration. However, specific devices need to be installed in the dwelling environment. Moreover, false alarms are often incurred by other sources causing the same effect. For example, an object fall might also cause similar pattern changes in vibration or sound. *Wearable sensor based fall detection systems* attempt to detect falls leveraging sensors embedded in wearable objects such as coat, belt and watch. The widely used sensors include accelerators, gyroscopes and barometric pressure sensors [6]. These detection systems can only work on the premise that all the devices are worn or carried by the subject during fall. *Smartphone based fall detector* is one of the promising fall detection systems with great potential due to the popularity of sensor-rich smartphones [7]. While these solutions are appropriate for fall detection in outdoor environment, the "always-on-the-body" requirements make the subject difficult to comply with, especially for the elders at home.

**Related Work on WiFi RF-based Activity Recognition.** The WiFi signal strength RSS has been exploited for indoor localization for more than a decade. However, only recently research attempts have been made to use WiFi RF signal for gesture and activity recognition [11][12]. While [11] first explores and uses WiFi RF signal to recognize different body or hand gestures, they use special instruments to collect special RF signals, which are not accessible with commodity WiFi devices. With the CSI tractable on commodity WiFi devices [9], Wi-Sleep [14] extracted the human respiration times in a controlled setting. And [10][13] employ RSS and CSI information respectively to detect the human motion in indoor environment.

The only work using WiFi commodity devices to detect fall is presented in [12], where it also exploits the WiFi CSI information at the granularity of OFDM subcarrier for fall detection. But the work only makes use of **the amplitude information of CSI** and differentiates the fall from few other specified activities. As all the human activities will cause variation in the amplitude of CSI across different sub-carriers, thus using the amplitude alone can only be used when a few human activities are specified. *When the elders live normally in the home environment with various activities, the solution will fail and produce huge number of false alarms*. In this paper, we intend to leverage both the amplitude and phase information of CSI from commodity WiFi devices to detect fall in real-time. Most importantly, we exploit **the phase difference over two antennas as the salient feature**, which was not explored before, to **robustly segment the fall and fall-like activities from the other activities**. Then with only the fall and few fall-like activities sifted out, we further employ both the amplitude and phase information to extract proper features to separate the real fall from other activities, which makes the real-time fall detection using the WiFi RF signal streams feasible in real home setting.

## 3 Preliminaries

In this section, we first introduce the Channel State Information (CSI) available on commodity WiFi devices, then define the fall activity types we aim to detect at home.

### 3.1 Channel State Information in 802.11n/ac

In frequency domain, the narrow-band flat-fading channel with multiple transmit and receive antennas (MIMO) can be modeled as

$$y = Hx + n$$

where $y$ and $x$ are the received and the transmitted signal vectors respectively, $n$ denotes the channel noise vector and $H$ is the channel matrix. Current WiFi standards (e.g., IEEE 802.11n/ac) use orthogonal frequency division modulation (OFDM) in their physical layer. OFDM splits its spectrum band (20MHz) into multiple (56) frequency sub-bands, called *subcarriers*, and sends the digital bits through these subcarriers in parallel. To estimate the channel matrix $H$, a known training sequence called the *pilot sequence* is also transmitted and the channel matrix $H$ is measured at the receiver side in the format of Channel State Information (CSI), which reveals a set of channel measurements depicting the amplitude and phase of every OFDM subcarrier. CSI of a single subcarrier is in the following mathematical format:

$$h = |h|e^{j\,sin\theta} \qquad (1)$$

where $|h|$ and $\theta$ are the amplitude and phase, respectively.

In indoor environments, WiFi signals propagate through multiple paths such as roof, floor, wall and furniture. If a person presents in the room, additional signal paths are introduced by the scattering of human body. WiFi RF-based activity recognition leverages the fact that *human activities cause the channel distortion, involving both amplitude attenuation and phase shift in the CSI streams*.

### 3.2 Fall Activity Types Targeted

There are many ways in which an elder can fall, and in this work we aim to detect the fall occurred in situations with respect to two transition activities: 1) Stand-fall refers to the situation that the fall occurs when an elder transfers out of a bed or chair, e.g., the elder may just stand up from the chair and feel dizzy due to cerebral ischemia; 2) Walk-fall refers to situation that the fall occurs while an elder is walking. According to a study by SignalQuest on falls in the elderly, 24% of falls occurred in the first case and 39% occurred in the second [19]. Hence, we aim for 63% of the fall situations in this work and plan to address the other fall types which occur while ascending or descending stairs or engaging in outdoor activities, in future work.

## 4 The Anti-Fall Fall Detection System

Our proposed real-time and non-invasive fall detector, Anti-Fall, consists of three functional modules: *the signal preprocessing module, the signal segmentation module and the fall detection module*. As shown in Figure 1, the system takes the CSI signal streams as input, which can be collected at the receiver side with commodity WiFi device (e.g., Intel 5300 NIC). The CSI signal streams are collected from each subcarrier (e.g., totally 30 subcarriers with Intel 5300 NIC) on a wireless link and totally two links are set up in the experiment between two antennas at the receiver side and one at the transmitter side. In order to obtain reasonably stable wireless signal for fall detection, each CSI signal stream is first preprocessed using a 1-D linear interpolation algorithm as suggested in [15], to ensure the received CSI evenly distributed in time domain. And the interpolated CSI signal stream is further processed by filtering out the temporal variations and long term changes, using a low-pass filter as suggested in [16]. After signal preprocessing, the CSI signal stream is fed into the core modules of Anti-Fall, which are *Activity Segmentation* and *Fall Detection* as shown in Figure 1.

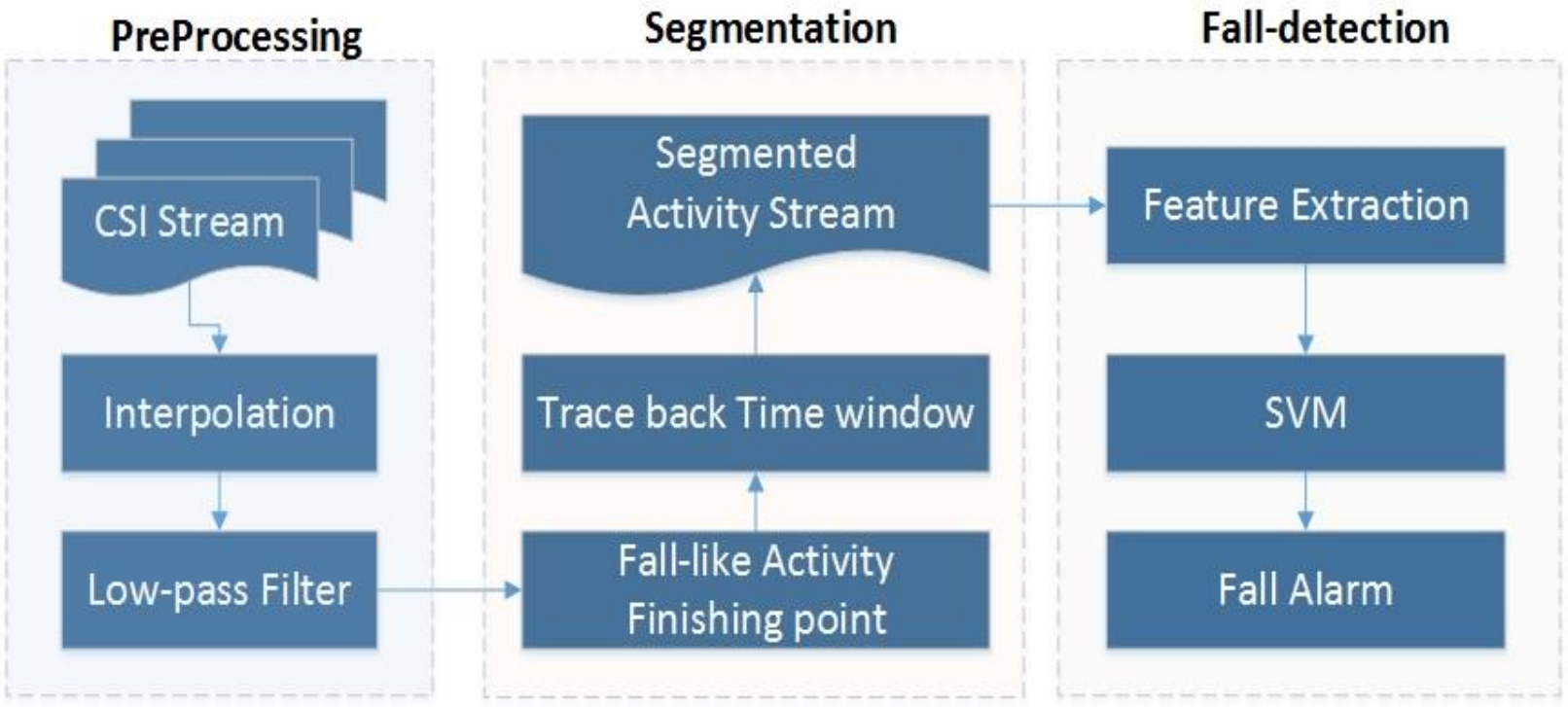

**Fig. 1.** Overview of the Fall Detection System Anti-Fall

### 4.1 Activity Segmentation

The main function of the activity segmentation module is to single out the fall and fall-like activities for further classification. It consists of two steps: in step 1 the ending point of the fall or fall-like activities is identified automatically by detecting the variance of CSI phase difference; then in step 2 the starting point of the fall or fall-like activities is determined by estimating the optimal window size of fall activities.

**Identify the ending point of fall or fall-like activities.** Through intensive experiments, we find that the variance of CSI phase difference over a pair of antennas is a very good and reliable indicator of human activities. Figure 2 shows the CSI phase difference variance of nine different human activities. Interestingly, it is observed that only several immobile human postures, such as *sitting still* and *lying*, result in very steady and stable signal patterns over the time. While most of the human activities, such as *walking*, *running*, *standing* and *falling*, all lead to obvious CSI phase difference fluctuation over time. Therefore, when people fall down, lie down or sit down, the variance of the CSI phase difference exhibits an obvious *transition from the fluctuation state to stable one*. To validate this observation, we recruit more people to conduct various daily activities, such as sweeping the floor, picking up objects, opening the window, etc. Through those experiments, *it is found that the state transition of the CSI phase difference variance is a robust feature to detect the fall and few other non-fall activities (i.e. lying down, sitting down).* We refer those few human activities which also result in obvious transition from fluctuation to stable state as *fall-like activities*. In this work, by using the state transition of CSI phase difference as the salient feature, we can automatically identify all the fall and fall-like activities in the continuously captured WiFi wireless signal streams.

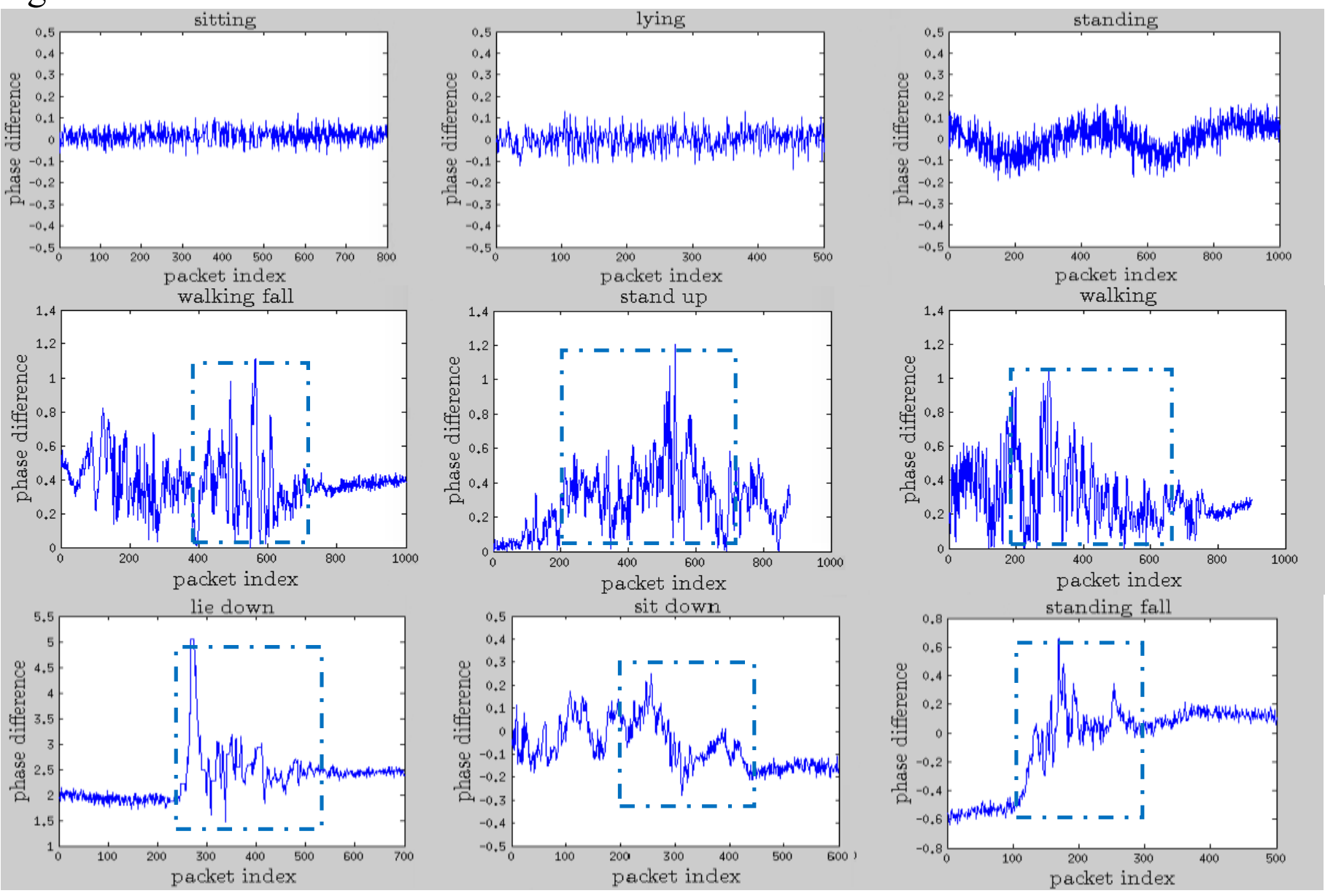

**Fig. 2.** Phase Difference Variance of Different Activities

In order to detect the ending point of fall and fall-like activities robustly, we need to detect the state transition of CSI phase difference by quantifying the stable state. We propose a *threshold-based sliding window method* to detect the stable state. First, we calculate the mean μ and the normalized standard deviation σ of CSI phase difference stream in stable state (e.g., *lying*) across multiple sliding windows off-line. Then, we compute the threshold value $\delta_{\text{threshold}}$ as follows:

$$\mu(V_{stable}) + 6\sigma(V_{stable}) <= \delta_{\text{threshold}}$$

Finally, we acquire the CSI phase difference variance in a sliding window and see if the whole sliding window lies in the stable state, by comparing the mean of CSI phase difference in the sliding window with the threshold $\delta_{threshold}$. Figure 3 shows the fall and fall-like human activity ending point identification results based on the state transition detection. It can be seen that *only the fall and fall-like activities are identified*, while other activities such as standing up and walking are left out.

**Determine the best window size for fall detection.** Based on the CSI phase difference state transition detection, we can identify the ending point of fall and fall-like activities in the continuously captured WiFi signal streams. To differentiate the fall from fall-like activities, we need to find the best window size to capture the fall and fall-like activities for accurate fall detection. In this regard, we propose a *two-phase approach* to search the optimal window size. First, we change the window size with a large step and evaluation the fall detection performance. After identifying a "good" window size range, we conduct a finer window size search and compare the fall detection performance, and in the end we choose the optimal window size based on the training dataset. In the evaluation section, we will report the window size search result.

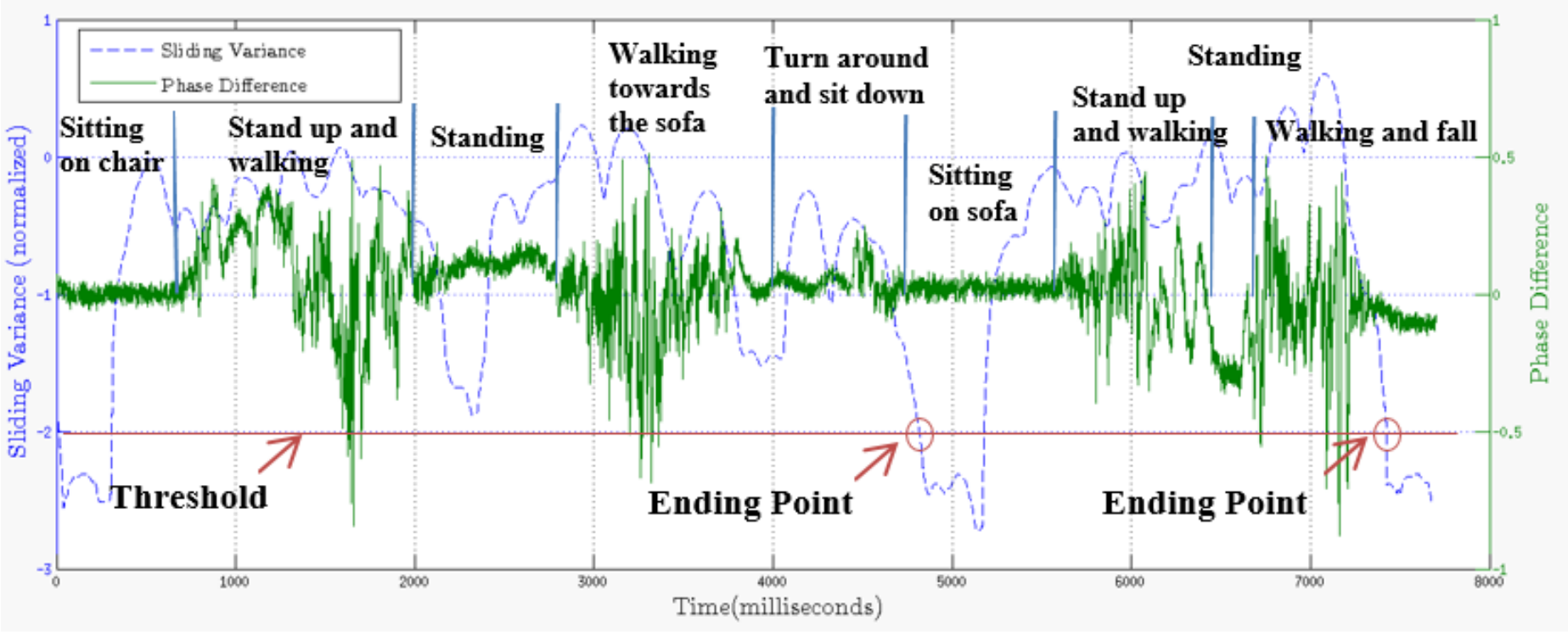

**Fig. 3.** Fall and Fall-like Activity Ending Point Identification ($\delta_{threshold}$ = -2)

## 4.2 Fall Detection

After determining the starting point and ending point in the *Activity Segmentation* module, only the CSI phase and amplitude of fall and fall-like activities are singled out. The goal of *Fall Detection* module is to separate the fall from fall-like activities.

**Feature Extraction.** We choose the following seven features as [12] for activity classification: (1) the normalized standard deviation (STD) of CSI, (2) the median absolute deviation (MAD), (3) the period of the activity, (4) the offset of signal strength, (5) interquartile range (IR), (6) signal entropy, (7) the velocity of signal change. However, different from [12] that only extracts features from the CSI amplitude information, we extract each of the above features from both CSI amplitude and phase information. Furthermore, since human activities affect different wireless links independently whereas affect neighboring subcarriers in a similar way [12], we select four subcarriers that spread evenly among all available 30 subcarriers. So each link generates the above seven features on amplitude and phase information in four subcarriers respectively and they together constitute the input of the SVM Classifier.

**SVM Classifier.** To detect the fall activity, a one-class Support Vector Machine (SVM) [17] is applied using the features extracted above. In one-class SVM, all the samples are divided into objective class (i.e., the fall) and non-objective class (i.e., fall-like activities). To solve the non-linear classification problem, it maps input samples into a high dimensional feature space by using a kernel function and find the maximum margin hyperplane in the transformed feature space. SVM classifier requires a training dataset and test dataset. In the process of classification model construction, fall and fall-like activities are segmented in the continuously captured WiFi wireless signal streams in the activity segmentation phase. Then the extracted features along with the corresponding labels are injected into the SVM classifier to build the classification model. In the process of real-time fall detection, the classification results along with the data will be recorded. With the user feedback, the wrong classification results will be re-labeled correctly and the model updating process will be triggered in time to update the classification model. We build the classification model by utilizing LibSVM [18].

## 5 Evaluation

In this section, we present the evaluation results of our Anti-Fall system using off-the-shelf WiFi devices. First, we introduce the experiment settings and the dataset. Second, we present the baseline method and metrics for evaluating Anti-fall briefly. Third, we report how fall detection results are affected by the activity window size. Then, the detailed evaluation results of Anti-Fall with respect to the baseline method are presented and compared. Finally, we show the system robustness with respect to environment changes.

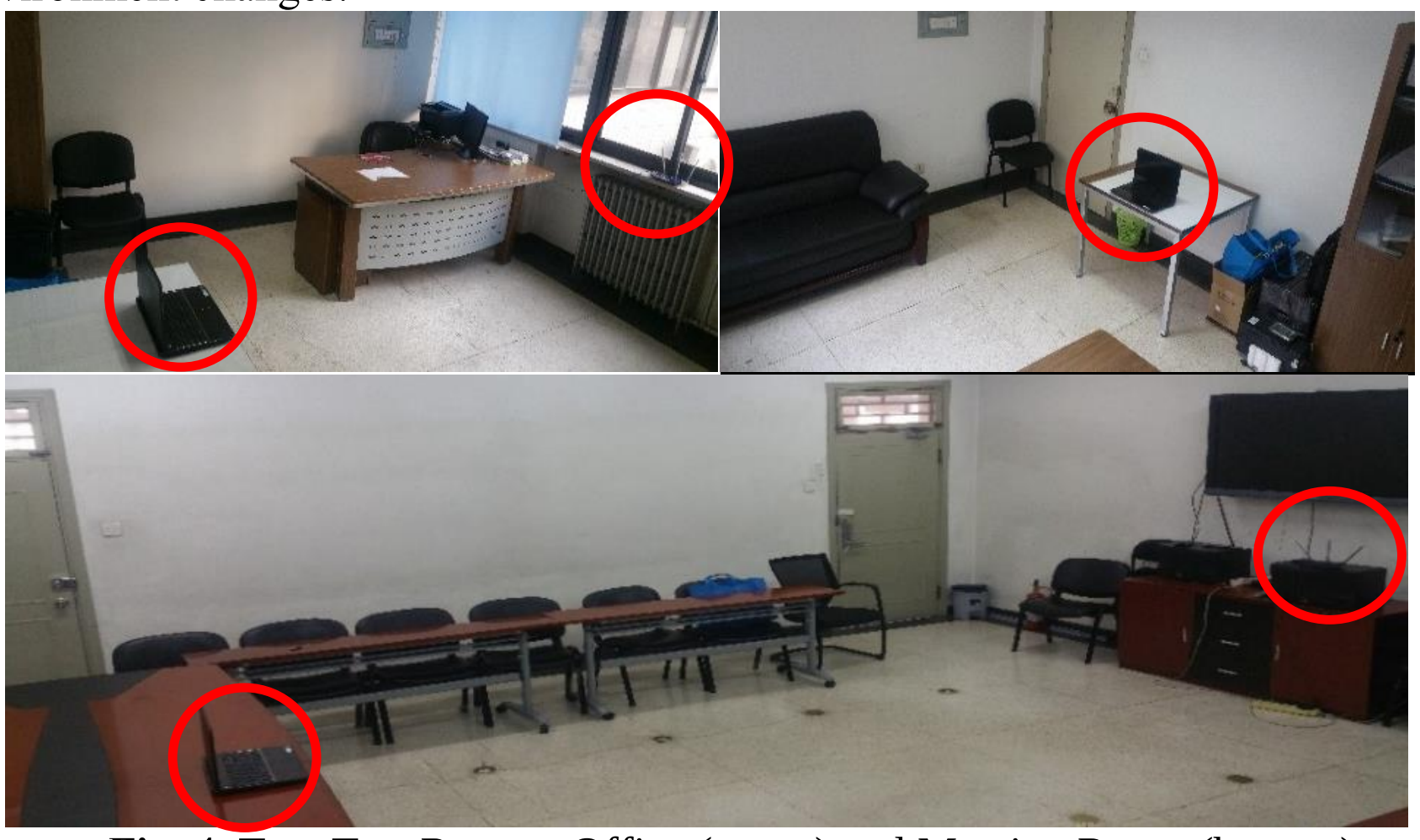

**Fig. 4.** Two Test Rooms: Office (upper) and Meeting Room (bottom)

## 5.1 Experimental Setups

We conduct experiments using an 802.11n WiFi network with one off-the-shelf WiFi device (i.e., a dell laptop with two internal antennas) connected to a commercial wireless access point (i.e., TP-Link WDR5300 Router with one antenna running on 5GHz). The laptop is equipped with an Intel WiFi Link 5300 card for measuring CSI [9]. The wireless data transmission rate is set to 100 packets per second.

We conduct experiments in two rooms of different size to test the generality of our approach. The experimental setups in these two rooms are shown in Figure 4. The smaller room (i.e. office) has the size of about 3m × 4m, whereas the larger one (i.e. meeting room) is about 6m × 6m.

## 5.2 Dataset

We recruit five male and one female students to perform various daily activities in the two test rooms over two weeks. Each data record consists of a few continuous activities, mixing the fall, fall-like and other activities. We deploy a camera in each room to record the activities conducted as ground truth. Over the test days, the chairs were moved to different places and the items on tables, such as bottles and bags, were moved, as usually occurs in daily life. During the experiments, the door of the room kept closed, and there was no furniture movement. The collected data records are processed by our transition-based segmentation method. We label the segmentation results according to the video records, finding that all 230 fall activities (70 in the meeting room and 150 in the office room) and 510 fall-like activities (160 in the meeting room and 350 in the office room) are all segmented correctly.

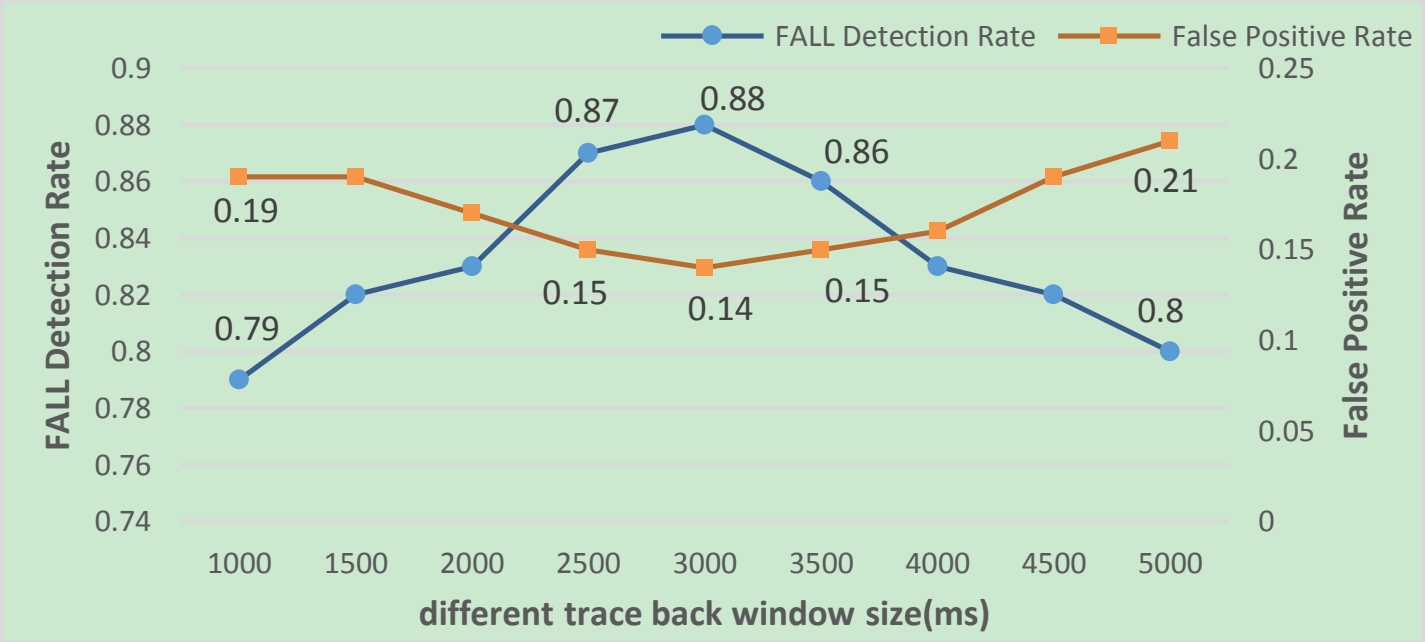

**Fig. 5.** Performance vs. Window Size (Coarse Search from 1s to 5s)

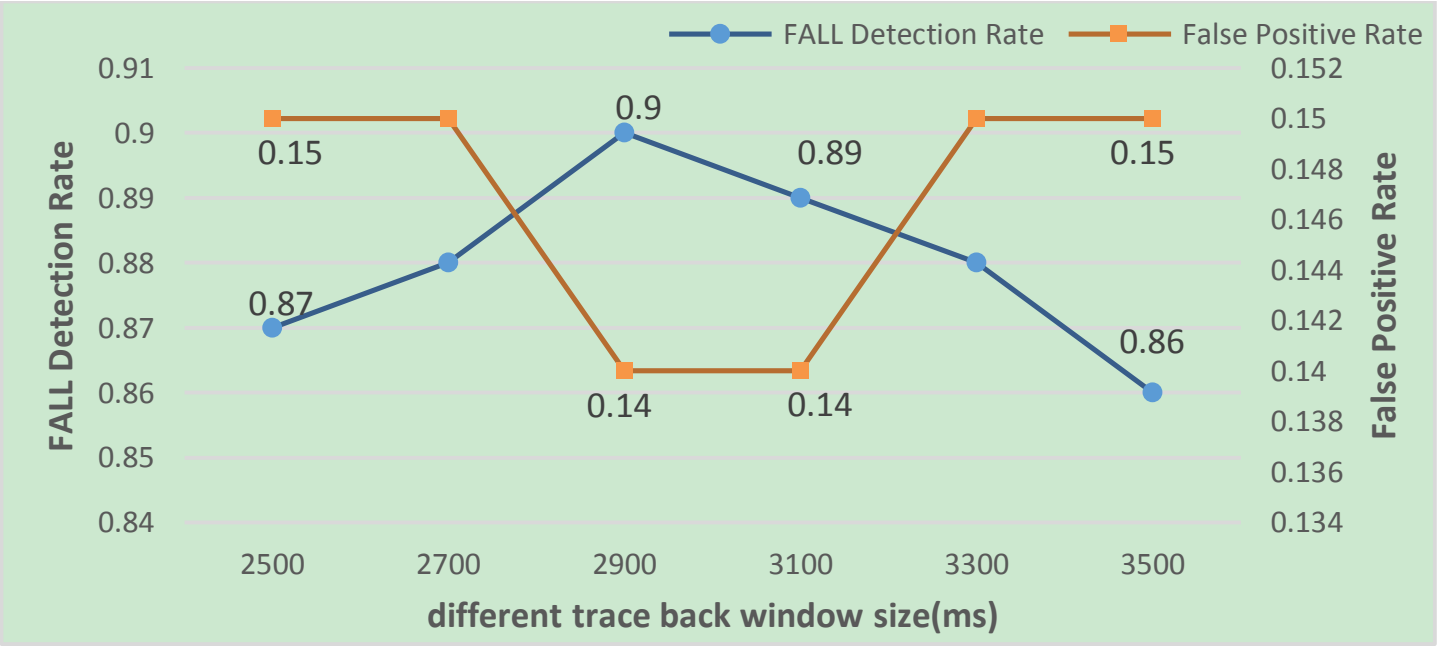

**Fig. 6.** Performance vs. Window Size (Fine Search from 2.5s to 3.5s)

### 5.3 Baseline Method and Metrics

In the experiments, we use the state-of-art fall detector WiFall proposed in [12] as the baseline. Since WiFall cannot segment the fall and other daily activities reliably, we thus leverage our proposed method to segment the fall and fall-like activities, subsequently we compare its activity classification method with our approach on our dataset. We use the following two standard metrics for performance comparison: **FALL Detection Rate (FDR)** indicates the proportion that the system can detect a fall correctly (true-positive). **False Positive Rate (FPR)** is defined as the proportion that the system generates a fall alarm when there is no fall happening.

### 5.4 Fall Detection Performance vs. Activity Window Size

Before we compare the fall detection performance of Anti-Fall with that of WiFall [12], we need to select the best activity window size using the method proposed in Section 4.1. We first use the dataset collected in the office room to show the relationship between the performance and window size. As shown in Figure 5, the best window size is between 2500ms to 3500ms using the coarse search method. The best window size is found to be 2900ms when a fine search is conducted between 2500ms and 3500ms, as shown in Figure 6. Then we use the dataset in the meeting room to repeat the same experiments, interestingly we get very similar optimal window size (3000ms). Thus, in all the evaluations we choose 3000ms as the test activity window size.

### 5.5 System Performance

In this part, we first compare the performance of Anti-Fall with that of the baseline method WiFall in terms of FDR and FPR. Then, we evaluate the system robustness of Anti-Fall system with respect to various environment changes.

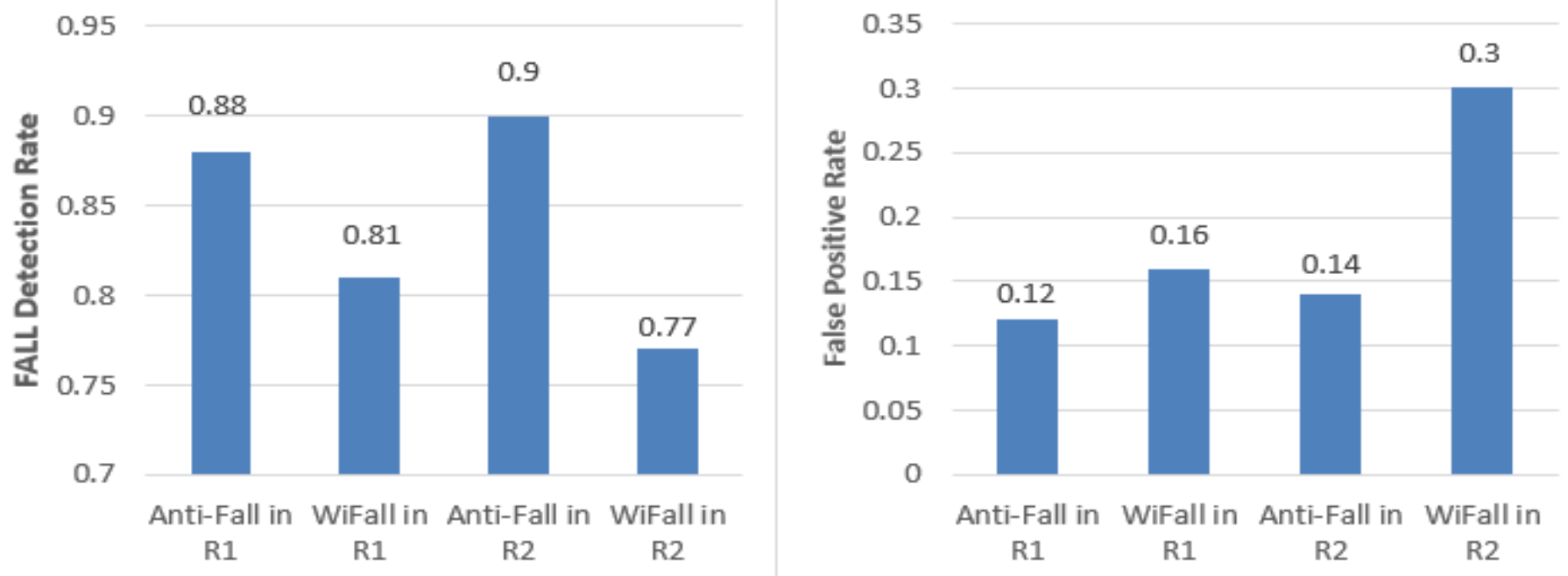

**Fig. 7.** FDR and FPR Results of Anti-Fall and WiFall in Two Test Rooms (R1 is the meeting room, R2 is the office room)

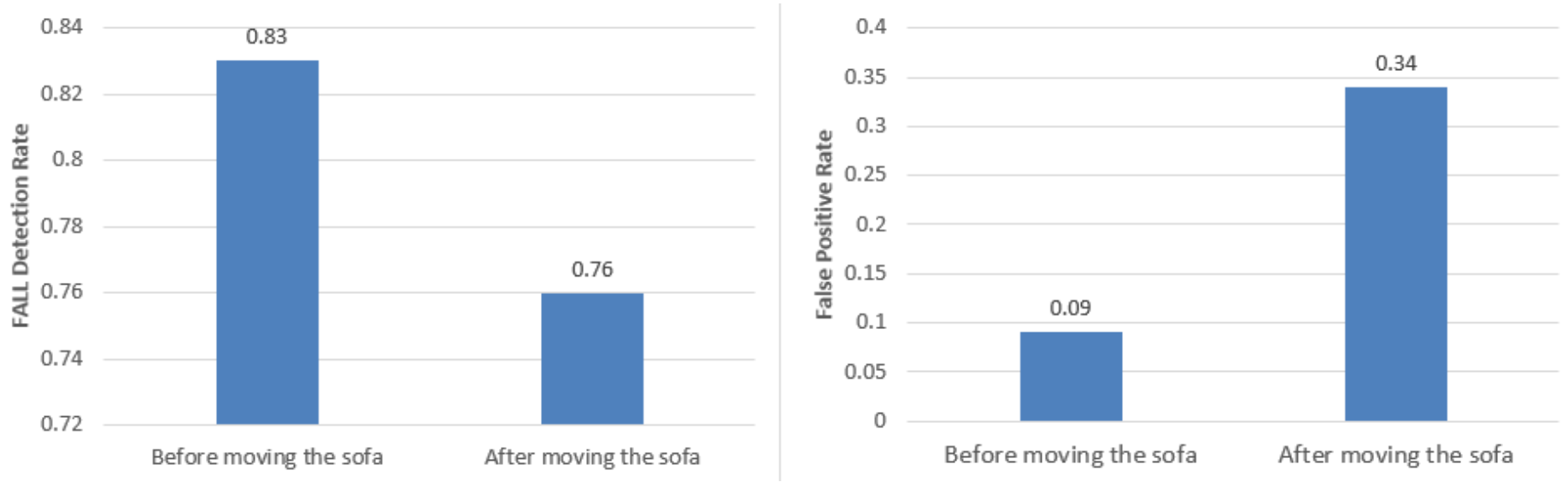

**Fig. 8.** Fall Detection Performance vs. Furniture (Sofa) Move

**Performance Comparison.** Figure 7 shows the performance of Anti-Fall with respect to the baseline method. Averaging the experimental results in both rooms, Anti-Fall achieves 89% detection rate and 13% false alarm rate. Compared to the baseline method WiFall, Anti-Fall gets 10% higher detection rate and 10% less false alarm rate.

**Robustness to Environment Changes.** As wireless signal is said to be very sensitive to environment changes, we thus evaluate the robustness of our approach against several common setting changes, including *opening the door and window, switching on/off the light, moving the furniture around*. While the Anti-Fall system performance is not affected much by *the opening of windows/door* or *the light on/off* in the two test rooms, its performance deteriorates when the furniture, such as the sofa, is moved. Specifically, when the sofa is moved from the window side to the door side in the office room as shown in Figure 8, the FDR drops from 83% to 76% while the FPR increases from 9% to 34%. So it seems that the furniture movement has quite a big impact on the fall detection performance, as it leads to significant CSI change due to wireless signal propagation path change, which requests classification model re-construction.

## 6 Conclusion

The availability of the Channel State Information (CSI) and multi-antenna capability in commodity WiFi devices has opened up new opportunities for activity recognition in recent years. In this work, we design and implement a non-intrusive, real-time and low-cost indoor fall detection system, called Anti-Fall, exploiting both the phase and amplitude information of CSI. To the best of our knowledge, **this is the first work** *to identify the CSI phase difference over two antennas as the salient feature to segment the fall and fall-like activities* reliably and *exploit both the phase and amplitude information of CSI for fall detection*. We have conducted extensive experiments and the evaluation results show that Anti-Fall is a very promising fall detection approach.

Fall detection has long been a research challenge in the public healthcare domain, especially for the elders. Although we implemented quite an effective fall detector using off-the-shelf WiFi devices, there are still many interesting problems that deserve further study. For example, can we apply the Anti-Fall solution in a multi-room home setting? How it works with the elders in real home setting where there are very few fall training data samples? Can we develop a very accurate personalized fall detector for an individual elder? How can we develop a fall detector which can adapt and evolve according to the environment change? We are working on these questions and expect to obtain promising results in near future.

## 7 Acknowledgment

This work is funded by the National High Technology Research and Development Program of China (863) under Grant No. 2013AA01A605. We would like to thank Wang Yibo, Wang Yuxiang, Li Xiang and Wu Dan for their help with experiments.

## 8 References

[1] Lord S.R., Sherrington C., Menz H.B., Close J.C.T.Falls in Older Peolpe: Risk Factors and Strategies for Prevention. Cambridge University Press; New York, NY, USA: 2001.

[2]CDC, "Falls among older adults: An overview," http://www.cdc.gov/HomeandRecreationalSafety/Falls/adultfalls.html, Apr. 2013.

[3] M. Alwan, P. J. Rajendran, S. Kell, D. Mack, S. Dalal, M. Wolfe, and R. Felder, "A smart and passive floor-vibration based fall detector for elderly," in *Information and Communication Technologies, 2006. ICTTA' 06. 2nd*, vol. 1. IEEE, 2006, pp. 1003–1007.

[4] Y. Li, K. Ho, and M. Popescu, "A microphone array system for automatic fall detection," *Biomedical Engineering, IEEE Transactions on*, vol. 59, no. 5, pp. 1291–1301, 2012.

*International Conference of the IEEE*. IEEE, 2007, pp. 1663–1666.

[5] X. Yu, "Approaches and principles of fall detection for elderly and patient," in *e-health Networking, Applications and Services, 2008. HealthCom 2008. 10th International Conference on*. IEEE, 2008, pp. 42–47.

[6] F. Bianchi, S. J. Redmond, M. R. Narayanan, S. Cerutti, and N. H. Lovell, "Barometric pressure and triaxial accelerometry-based falls event detection," *Neural Systems and Rehabilitation Engineering, IEEE*

*Transactions on*, vol. 18, no. 6, pp. 619–627, 2010.

[7] J. Dai, X. Bai, Z. Yang, Z. Shen, and D. Xuan, "Perfalld: A pervasive fall detection system using mobile phones," in *Pervasive Computing and Communications Workshops (PERCOM Workshops), 2010 8th IEEE International Conference on*. IEEE, 2010, pp. 292–297.

[8] P. Natthapon, S. Thiemjarus, and E. Nantajeewarawat. "Automatic Fall Monitoring: A Review." *Sensors, July* 2014, 12900–12936.

[9] D. Halperin, W. Hu, A. Sheth, and D. Wetherall, "Predictable 802.11 packet delivery from wireless channel measurements," SIGCOMM Comput. Commun. Rev., vol. 40, no. 4, pp. 159–170, Aug. 2010.

[10] A. E. Kosba, A. Saeed, and M. Youssef, "Rasid: A robust wlan devicefree passive motion detection system," in Proceedings of IEEE PerCom. IEEE, 2012, pp. 180–189.

[11] Q. Pu and et al., "Whole-home gesture recognition using wireless signals," in *ACM MobiCom*, pp. 27–38, 2013.

[12] C. Han, K. Wu, Y. Wang, and L. Ni, WiFall: "Device-free fall detection by wireless networks", in Proc. of 33rd IEEE Int. Conf. on Computer Communications, Toronto, Canada, 2014, pp. 271–279.

[13] Kun Qian, Chenshu Wu, Zheng Yang, Yunhao Liu, Zimu Zhou,"PADS: Passive Detection of Moving Targets with Dynamic Speed using PHY Layer Information", IEEE ICPADS, Hsinchu, Taiwan, December 16 - 19, 2014.

[14] X. Liu, J. Cao, S. Tang, and J. Wen, “Wi-Sleep: Contactless Sleep Monitoring via WiFi Signals,” in IEEE RTSS, 2014.

[15] R. Nandakumar, B. Kellogg, S. Gollakota, “Wi-Fi Gesture Recognition on Existing Devices,” arXiv:1411.5394v1

[16] Wang, Yan, et al. "E-eyes: device-free location-oriented activity identification using fine-grained WiFi signatures." Proceedings of the 20th annual international conference on Mobile computing and networking. ACM, 2014.

[17] B. Scholkopf, J. C. Platt, J. Shawe-Taylor, A. J. Smola, and R. C. Williamson, “Estimating the support of a high-dimensional distribution,” Neural computation, vol. 13, no. 7, pp. 1443–1471, 2001.

[18] C.-C. Chang and C.-J. Lin, “Libsvm: a library for support vector machines,” ACM Transactions on Intelligent Systems and Technology (TIST), vol. 2, no. 3, p. 27, 2011.

[19] J. Shea, An Investigation of Falls in the Elderly, Available from URL: http://www.signalquest.com/master%20frameset.html?undefined, July 2005